\documentclass[aps,prb,preprint,superscriptaddress,groupeaddress]{revtex4}
\usepackage{graphicx}
\usepackage{amsmath}
\usepackage[Symbol]{upgreek}

\begin{document}

\title{Coherent Charge Transport in Ballistic InSb Nanowire Josephson Junctions}

    \author {S. Li}
    \affiliation{Key Laboratory for the Physics and Chemistry of Nanodevices and Department of Electronics, Peking University, Beijing 100871, China}
    \author {N. Kang}
    \email[Corresponding author: ]{nkang@pku.edu.cn}
    \affiliation{Key Laboratory for the Physics and Chemistry of Nanodevices and Department of Electronics, Peking University, Beijing 100871, China}
    \author {D. X. Fan}
    \affiliation{Key Laboratory for the Physics and Chemistry of Nanodevices and Department of Electronics, Peking University, Beijing 100871, China}
    \author {L. B. Wang}
    \affiliation{Key Laboratory for the Physics and Chemistry of Nanodevices and Department of Electronics, Peking University, Beijing 100871, China}
    \author {Y. Q. Huang}
    \affiliation{Key Laboratory for the Physics and Chemistry of Nanodevices and Department of Electronics, Peking University, Beijing 100871, China}
    \author {P. Caroff $^{a)}$\footnotemark[0]\footnotetext[0]{$^{a)}$Present address: Department of Electronic Materials Engineering, Research School of Physics and Engineering, The Australian National University, Canberra, ACT 0200, Australia}}
    \affiliation{I.E.M.N., UMR CNRS 8520, Avenue Poincar\'e, BP 60069, F-59652 Villeneuve d'Ascq, France}
    \author {H. Q. Xu}
    \email[Corresponding author: ]{hqxu@pku.edu.cn}
    \affiliation{Key Laboratory for the Physics and Chemistry of Nanodevices and Department of Electronics, Peking University, Beijing 100871, China}
    \affiliation{Division of Solid State Physics, Lund University, Box 118, S-221 00 Lund, Sweden}
    \date{\today}

\begin{abstract}

Hybrid InSb nanowire-superconductor devices are promising for investigating Majorana modes and topological quantum computation in solid-state devices.
An experimental realisation of ballistic, phase-coherent superconductor-nanowire hybrid devices is a necessary step towards engineering topological superconducting electronics. Here, we report on a low-temperature transport study of Josephson junction devices fabricated from InSb nanowires grown by molecular-beam epitaxy and provide a clear evidence for phase-coherent, ballistic charge transport through the nanowires in the junctions. We demonstrate that our devices show gate-tunable proximity-induced supercurrent and clear signatures of multiple Andreev reflections in the differential conductance, indicating phase-coherent transport within the junctions. We also observe periodic modulations of the critical current that can be associated with the Fabry-P\'{e}rot interference in the nanowires in the ballistic transport regime. Our work shows that the InSb nanowires grown by molecular-beam epitaxy are of excellent material quality and hybrid superconducting devices made from these nanowires are highly desirable for investigation of the novel physics in topological states of matter and for applications in topological quantum electronics.

\end{abstract}

\maketitle


Study of coherent charge transport through superconductor-normal conductor-superconductor (SNS) Josephson junction devices is of fundamental importance for understanding the properties of superconducting quantum devices\cite{SNS, Yeyati}.
When a normal conductor is connected to two closely spaced superconductor leads, Cooper pairs can leak from the superconductor leads into the normal conductor, due to the superconducting proximity effect, and induce a supercurrent through the junction.
The advances in nanofabrication technology and material science have made it possible to experimentally realise a variety of Josephson junctions with reduced dimensions.
These mesoscopic superconducting hybrid structures realisable with various types of materials have offered great possibilities to investigate novel quantum phenomena and new device concepts including, e.g., tunable supercurrent transistors\cite{Doh, Xiang, Jarillo, SiGe, C60}, Andreev bound states\cite{CEA, Lee}, Cooper pairs splitters\cite{Basel, Strunk}, superconducting electronic refrigerators\cite{Giazotto}, and topological superconducting states of matter\cite{Kane, Lutchyn01, Oreg}.
Recently, Josephson junctions based on InSb nanowires have been attracting growing attention owing to their extraordinary transport properties, such as long coherence length, tunable one-dimensional subbands, good contacts with superconductor metals, and strong spin-orbit coupling\cite{Nilsson, Plissard2013, Mourik, Deng}.
In particular, signatures of Majorana zero-energy states have been observed in hybrid InSb nanowire-superconductor devices\cite{Mourik, Deng, Churchill}. Beside their interest in fundamental physics, Majorana bound states in solid state systems have potential application in topological quantum computation\cite{Alicea, Heck, Sau}.
Moreover, the physics of the Josephson junctions made from InSb nanowires is rich due to the interplay between strong spin-orbit coupling and Zeeman splitting in the presence of a magnetic field. For example, various novel physical phenomena, such as the fractional Josephson effect, anomalous Josephson current and bound subgap states\cite{Fu, Cheng, Nazarov1, Cayao}, have been predicted to be observable in such hybrid quantum devices.
Therefore, it is of broad interest and fundamental importance to study the proximity-induced superconducting properties of InSb nanowire-based hybrid structures.

So far, most of the experimental works on InSb nanowire-superconductor junctions have been carried out in the diffusive transport regime\cite{Nilsson, Mourik, Deng}.
The combined effects of phase-coherent transport and proximity-induced superconductivity in the ballistic transport regime have yet to be systematically explored in InSb nanowire-based Josephson junctions.
There are two major challenges in the realisation of a ballistic InSb nanowire-based Josephson junction.
One is to achieve highly transparent interfaces between superconductor contacts and an InSb nanowire, and the other is to obtain an InSb nanowire with perfect crystal quality. While the first one is essential to enhance the proximity-induced superconductivity in the nanowire junction, the second one is critical for achieving a long mean free path, desired for ballistic transport, in the nanowire.
Recently, we have succeeded in growing high crystal quality InSb nanowires by molecular-beam epitaxy (MBE)\cite{XuNanotech2012, Thelander2012}.
The cleanliness of the MBE growth environment enables growth of semiconductor nanowires with high purity and perfect crystal quality.
A transport study of MBE-grown InSb nanowires has demonstrated the formation of long single quantum dots with lengths up to 700 nm\cite{Fan}. Considering the fact that an alternative theoretical scenario, taking into account the effect of disorder, has been put forward to explain the previously observed zero-energy states\cite{Liu}, these state-of-the-art InSb nanowires are desirable for exclusion of such explanations in Majorana fermion detection experiments and for the study of novel ballistic, quantum coherent transport phenomena.

In this work, we report on realisation and systematic low-temperature transport measurements of superconductor-InSb nanowire-superconductor Josephson junction devices with highly transparent contacts to superconductor leads, in which the InSb nanowires are grown by MBE. This is the first report on a systematic study of the transport properties of MBE-grown InSb nanowire Josephson junction devices. We demonstrate that the devices show proximity-induced supercurrent and clear signatures of multiple Andreev reflections, indicating that the charge transport is phase-coherent within the junctions. We also observe periodic modulations of the critical current and show that the modulations are associated with the resonant states formed by Fabry-P\'{e}rot interference in the nanowires. Our results demonstrate that transport in MBE-grown InSb nanowire Josephson junctions is ballistic and phase-coherent, and that such nanowires are appropriate for investigation of novel mesoscopic superconducting phenomena in the presence of strong spin-orbit interaction and for detection and manipulation of Majorana bound states in the solid state.

\noindent
\textbf{Results}

\noindent
\textbf{Josephson junctions based on individual InSb nanowires.}

\noindent
Figure 1a shows a scanning electron microscope (SEM) image of an array of as-grown vertically standing InAs/InSb heterostructure nanowires on an InP(111)B substrate. Single nanowire Josephson junction devices with aluminum electrodes are fabricated from the as-grown InSb nanowires on a highly doped silicon substrate (used as a global back gate) covered with a 200-nm-thick SiO$_{2}$ capping layer. The inset of Figure 1d displays a representative single nanowire Josephson junction device. The diameters of the InSb nanowires used in this work range from 50 to 120 nm and the lengths are in the range of 1-2 $\mu$m. The source-drain separations $L$ of the devices are in the range of 60-200 nm. Figure 1b shows the differential conductance $dI_{sd}/dV_{sd}$ measured for a device with $\sim$56 nm in nanowire diameter and $\sim$60 nm in contact separation (device D1) as a function of source-drain bias voltage $V_{sd}$ at base temperature $T$ = 10 mK, far below the superconducting critical temperature of aluminum ($T_{c}\approx$ 1.1 K). The conductance shows a pronounced peak at $V_{sd}=0$ V, indicating the occurrence of a supercurrent flowing through the nanowire. The conductance rises when the bias voltage is decreased below $V_{sd} = 2\Delta/e \approx 300$ $\upmu$eV , where $\Delta\approx 150$ $\upmu$eV and $2\Delta$ is the superconducting energy gap of aluminum. It is worth noting that the conductance at bias voltages lower than the superconducting energy gap is nearly doubled compared to the normal-state conductance. This is evidence that contact interfaces between the superconductor electrodes and the nanowire are highly transparent and is fully consistent with the Andreev reflection mechanism of transport through a Josephson junction with two almost ideal transparent contact interfaces as described by the Blonder-Tinkham-Klapwijk (BTK) theory\cite{BTK}. At finite bias voltages, subharmonic gap structures, symmetrically situated around $V_{sd}=0$ V, can be clearly observed in the measured conductance. Such  subharmonic gap structures are commonly observed in an SNS junction\cite{Xiang, Jarillo, Nilsson} and  could be understood by invoking multiple Andreev reflection (MAR) processes. In a MAR process, an electron is reflected multiple times, gaining an energy of $eV_{sd}$ in each reflection, until its energy is above the superconducting energy gap. This process generates a series of conductance peaks within the superconducting energy gap, with different peaks corresponding to different orders of successive Andreev reflections\cite{Elke, Cuevas}. As shown in the inset of Figure 1b, the voltage positions $V_{peak}$ of the differential conductance peaks fall on top of the expected linear dependence $V_{peak}=2\Delta/ne$, where $n$ is an integer number. Each data point in the inset is obtained from an average of several $dI_{sd}/dV_{sd}$ traces at different gate voltages and the error bar attached to the data point denotes the standard deviation. The evolutions of these subharmonic conductance peaks with temperature are shown in Figure 1c. The positions $V_{peak}$ of the differential conductance peaks show systematic shifts towards lower bias voltages with increasing temperature. At temperatures above the critical temperature $T_{c}$ of aluminum, these subharmonic gap structures disappear.
Also, as expected, the positions of the subharmonic gap structures follow the BCS-type temperature dependence, in which the superconducting energy gap varies with temperature approximately as $\Delta(T)\approx \Delta(0)\sqrt{\cos[\frac{\pi}{2}(\frac{T}{T_{c}})^2]}$ (see the dashed lines in Figure 1c for the evolutions of the peak positions for $V_1=2\Delta/e$ and $V_2=\Delta/e$). The observation of MARs implies the coherent transfer of multiple charges in our junction device.

The transparency of the interfaces between the InSb nanowire and the aluminum contacts can be estimated from the excess current $I_{exc}$---a non-zero current offset at $V_{sd}=0$ V as obtained by the linear extrapolation of the current-voltage characteristics at high bias voltages ($V_{sd} > 2\Delta/e$)---due to the transfer of Cooper pairs through the junction via Andreev reflection processes.
In the BTK theory for Andreev reflection, the strength of the potential barrier at a superconductor-normal conductor interface can be characterized by a dimensionless scattering parameter $Z$, which relates to the transmission coefficient at the interface as $T_{r}=1/(1+Z^{2})$\cite{Flensberg, OBTK}. The scattering parameter Z can be estimated from the ratio $eI_{exc}R_{n}/\Delta$, where the normal state resistance $R_{n}$ and $I_{exc}$ can be determined from the linear fit to the measured current-voltage curve at high bias voltages and its intercept with the current axis at zero bias voltage, respectively. The current-voltage characteristics of the junction measured at gate voltage $V_{g} = -17.7$ V and base temperature $T=10$ mK are shown in Figure 1d. Using $\Delta=150$ $\upmu$eV, we obtain $Z\sim 0.42$, which corresponds to $T_{r}=$ 0.85 at the interfaces, indicating that the nanowire-aluminium contacts are highly transparent with the transmission close to the ideal value of unity. To the best of our knowledge, this high value of the transparency is the best result ever achieved for InSb nanowire devices contacted by superconductor electrodes. We attribute this high transparency at the nanowire-superconductor interfaces to both the high material quality of the MBE-grown InSb nanowire and the improvement in the contact fabrication technique (see Methods).

In Figure 2a we show current-biased, four-probe, measurements of the voltage-current characteristics  of the device at different gate voltages $V_{g}$ and base temperature $T=10$ mK. A characteristic dc Josephson response is observed---each $V_{sd}-I_{bias}$ curve exhibits a zero-voltage region when the current $I_{bias}$ is below a certain value $I_{c}$, providing a clear evidence for the proximity-induced superconductivity in the hybrid superconductor-nanowire device. We can observe an induced supercurrent through a hybrid superconductor aluminium-InSb nanowire device with a junction length up to 200 nm (see Supplementary Fig. S1). Beyond $I_{c}$,  switching occurs from the superconducting state to the dissipative resistive state, leading to the abrupt appearance of a finite source-drain voltage. To show the dependence of $I_{c}$ on $V_{g}$ in more detail, we plot in the top panel of Figure 2b the differential resistance $dV_{sd}/dI_{bias}$ as a function of the applied gate voltage $V_g$ and bias current $I_{bias}$. In the central dark region, $dV_{sd}/dI_{bias}=0$ and the nanowire is in the superconducting state with the critical current $I_{c}$ given by the thin bright lines of the high resistance in the figure. Here it is easily seen that the magnitude of the critical current $I_{c}$ is gate voltage-dependent. In the areas outside the central dark region, the nanowire is in the dissipative state. The bottom panel of Figure 2b displays the evolution of the corresponding normal state conductance $G_{n}$ over the same $V_{g}$ range, where $G_{n}$ is determined from linear fits of the measured $I_{sd}-V_{sd}$ curves at high source-drain voltages ($V_{sd} > 2\Delta/e$). It can be readily seen that the critical current is closely correlated with the normal state conductance $G_{n}$. Such correlations between $I_{c}$ and $G_{n}$ have been established for SNS junctions both in the diffusive and in the ballistic transport regime\cite{Likharev, Beenakker02}.
Based on field-effect electrical measurements, we are able to determine an electron density of $n_{e}\simeq 9\times 10^{17} cm^{-3}$ and a mobility of $\mu\simeq$ 2000-4000 cm$^{2}$/V$\cdot$s in the nanowire. These values give an electron mean free path $l_{e}\sim$ 40-80 nm, which is comparable to the contact separation in the device. The superconducting coherence length is given by $\xi =(\hbar D/\Delta)^{1/2} \simeq$ 500 nm, where D is the diffusive coefficient. Therefore, our junction operates in the short junction ($L\leq \xi$) and ballistic or quasi-ballistic regime ($L\leq l_{e}$). According to theory, the product $I_{c}R_{n}$ of the critical current and normal state resistance in the clean short-junction limit should be $\pi\Delta/e\approx 470$ $\upmu$V.\cite{Beenakker02} As shown in Figure 2c, the measured values, $I_{c}R_{n}\approx 10-30$ $\upmu$V, are however much smaller than expected. Such strong suppression of the $I_{c}R_{n}$ product has been commonly observed in a Josephson junction device and has been attributed to premature switching in such a resistively and capacitively shunted Josephson junction device\cite{Doh, Xiang, Jarillo, Nilsson, Tinkham}.

\noindent
\textbf{Fabry-P\'{e}rot interference in ballistic InSb nanowire Josephson junctions.}

It is seen in Figure 2b that the modulations of $I_c$ are strongly correlated with the variations of the normal state conductance of the junction in the open conduction region where the normal state conductance lies in a range of 3 to 8 $e^{2}/h$. Such correlated $I_c$ modulations and $G_n$ variations have been previously observed in several semiconductor-based Josephson junctions\cite{Doh, Xiang, Nitta, Abay} and have been attributed to the gate voltage induced changes in the carrier density. Below, we will show further that the observed modulations of $I_{c}$ and variations of $G_{n}$ arise from the phase-coherent transport through the Fabry-P\'{e}rot-type resonant states formed in the clean nanowire junction region and show that, as observed in carbon nanotube-based ballistic SNS junctions\cite{Jarillo, Lindelof, Lau}, the critical current and the normal state conductance exhibit regular periodic oscillations.

Figure 3a displays a gray scale plot of the differential conductance $dI_{sd}/dV_{sd}$ for device D2, where the nanowire's diameter is $\sim$60 nm and the contact seperation is $\sim$95 nm, as a function of $V_{g}$ and $V_{sd}$. A small magnetic field of $B=$ 50 mT is applied perpendicularly to the device substrate in order to suppress superconductivity in the electrodes. The differential conductance exhibits a characteristic ``checkerboard'' pattern (see dashed lines in Figure 3a for a guided view), showing quasi-periodic oscillations with both $V_{sd}$ and $V_{g}$. The oscillations can be understood qualitatively by ballistic electron transport through the resonant states formed within the electron cavity confined by the two metal contacts in analogy to an optical Fabry-P\'{e}rot interferometer. In a ballistic device, the phase-coherent electron transport in the nanowire between the contacts can lead to transmission resonances when the Fermi wave vector $k_{F}$ of electrons in the nanowire matches the condition for resonance. By tuning either $V_{sd}$ or $V_{g}$, a shift in the Fermi energy of electrons in the nanowire leads to a change in the Fermi wave vector, giving rise to a shift in the condition for resonance along a crisscrossing mesh line. Though similar patterns of periodic conductance oscillations have previously been observed in short junction devices\cite{Liang, Javey, Cao, Kretinin, Miao, Hakonen}, there has not been a systematic report for observation of such patterns in a device made from an InSb nanowire. With a parabolic electron dispersion in a nanowire, the smallest energy spacing $\Delta E$ of Fabry-P\'{e}rot resonances can be expressed as $\Delta E=\hbar^{2}\pi^{2}/2m^{*}L_{c}^{2}$, where $\hbar$ is the Planck constant, $m^{*}$ is the effective mass and $L_{c}$ is the length of the resonant cavity\cite{Kretinin}. For device D2, the observed value of this energy spacing is $\Delta E\simeq 2-3$ meV, corresponding to an effective cavity length $L_c \approx 75-115$ nm, in good agreement with the geometrical length of the junction, i.e., the contact separation, in the device. The observation of a regularly oscillating conductance interference pattern implies that the charge transport through the nanowire in the junction is in the ballistic regime.

In Figure 3b we plot the critical current $I_{c}$, extracted from voltage-current characteristics of device D2 measured at zero magnetic field, over the same $V_{g}$ range as in Figure 3a. The critical current $I_{c}$ displays clearly quasi-periodic variations with the gate voltage $V_g$. In a nanowire electron Fabry-P\'{e}rot cavity, quantum interference leads to a set of resonant levels. In the case of the nanowire being in the proximity-induced superconducting state, the Josephson current can flow through these discretely resonant levels, giving rise to periodic modulations of $I_{c}$\cite{Glazman, Beenakker02, Rittenhouse}. As shown in Figure 3b, the critical current exhibits peaks at the same gate voltages where the device shows maximal normal state conductance values, providing a compelling evidence that such Fabry-P\'{e}rot constructive interference manifests itself in the superconducting states. Considering a Josephson junction in an electromagnetic environment, the critical current through quantized energy levels in the resonant regime is given by\cite{Jarillo} $I_{c}\sim[1-(1-\frac{1}{4}\frac{G_{n}}{G_{0}})^{1/2}]^{3/2}$. This correlation between $I_{c}$ and $G_{n}$ explains our observation shown in Figures 3a and 3b. In addition, we note that its deviation from the linear dependence of $I_c$ on $G_n$ predicted for an ideal SNS junction could explain the observed $V_{g}$ dependence of $I_{c}R_{n}$ as shown in Figure 2c.

\noindent
\textbf{Discussion}

As shown in Figure 3a, we can clearly identify two sets of periodic oscillations in the interference pattern, highlighted by dashed lines and dotted lines, respectively. As we will discuss below, these two sets of periodic oscillations originate from the multimode transport in the InSb nanowire. In our devices, the shape of the nanowires can be viewed as a cylinder where the diameter of the nanowire $D$ is comparable to the channel length $L$. As a result, both the transverse quantization and the longitudinal Fabry-P\'{e}rot interference contribute to the formation of the resonant states in the nanowires. The relative magnitudes of the energy quantization resulting from the transverse confinement and the longitudinal Fabry-P\'{e}rot interference can be approximately determined based on the aspect ratio $\xi \sim D/L$. In device D2 with $L\sim$ 95 nm and $D\sim$ 60 nm ($\xi > 1$), the fast oscillations with a period of $\sim$ 1.6 V in $V_{g}$ originate from the longitudinal Fabry-P\'{e}rot interference and the slow oscillations correspond to the transverse modes. Such resonant conductance and critical current oscillations are also observed in the other nanowire devices with different aspect ratios. In Figure 4, we display the differential conductance measured as a function of $V_{g}$ and $V_{sd}$ for device D3 with $L\sim$ 65 nm and $D\sim$ 90 nm ($\xi < 1$) at zero magnetic field. At low bias voltages ($V_{sd}\le 2\Delta/e$), the proximity-induced supercurrent exhibits quasi-periodic oscillations with $V_g$. At high bias voltages ($V_{sd} > 2\Delta/e$), we also observed two sets of Fabry-P\'{e}rot oscillations. As expected, the larger diameter of the nanowire results in smaller resonant energy level spacings and thus a shorter period of conductance oscillations. Similar complex interference patterns have been previously observed in the devices made from two-dimensional electron gas and graphene, and have been attributed to the two-dimensional geometry of the cavities\cite{Miao, Hakonen, Katine, White, Trauzettel, Zalipaev}. Finally, it is worthy to note that our devices operate in the high conductance regime, where more than one subbands are occupied in the nanowires, which could give rise to a beating behaviour in conductance oscillations\cite{Kretinin}.


In summary, we have investigated low temperature transport properties of Josephson junction devices made from high crystal quality, MBE-grown InSb nanowires with highly transparent interfaces between superconductor electrodes and the nanowires. We have observed that the devices show gate-tunable proximity-induced supercurrent and signatures of multiple Andreev reflections in the differential conductance, as expected for phase-coherent transport through the nanowire junctions. We have also observed that the devices exhibit characteristic Fabry-P\'{e}rot oscillations in the normal state conductance, a manifestation of ballistic transport through the nanowire junctions, and modulations of the supercurrent associated with the resonant states formed by Fabry-P\'{e}rot interference in the nanowires. This work presents a first report on a systematic study of ballistic InSb nanowire based Josephson junction devices and provides an important step towards development of topological quantum computation devices based on hybrid InSb nanowire-superconductor structures.

\noindent
\textbf{Methods}

\noindent
\textbf{InSb nanowire growth and device fabrication.}
Single crystalline InSb nanowires used in this study were grown by gas source MBE on semi-insulating InP(111)B substrates, using a known stem technique\cite{Thelander2012, CaroffSmall2008, ErcolaniNanotech2009, PlissardNL2012}. The grown InSb segments were confirmed to be 100\% perfectly untwined zinc blende single crystal nanowires, as usually observed for gold-assisted growth by other techniques\cite{CaroffSmall2008, ErcolaniNanotech2009}. All nanowires showed some lateral overgrowth but no measurable tapering. Details on the growth and characterization of the nanowires can be found elsewhere\cite{XuNanotech2012}.

After mechanical transfer from a growth chip to a Si/SiO$_2$ substrate prepatterned with contact pads and alignment markers on top of the SiO$_2$ layer, individual InSb nanowires were located using SEM. The superconducting electrodes consisting of 5-nm-thick titanium and 90-nm-thick aluminum were defined on the located individual InSb nanowires by electron-beam lithography (EBL), electron-beam evaporation of metal and lift-off processes. To lower the contact resistance between InSb nanowires and the metallic electrodes, the native oxide layers on the surfaces of the InSb nanowires were removed using an ammonium polysulfide ((NH$_{4}$)$_{2}$S$_{x}$) solution prior to metal deposition.\cite{Suyatin} Nanowire devices with resistance at room temperature in a range of 5-30 k$\Omega$ were selected for low-temperature transport measurements in this work.
Comparing with earlier works,\cite{Nilsson, Deng} we used longer time for etching and passivation treatment in order to achieve highly tranparent contact interfaces. Before the devices were cooled down for low-temperature transport measurements, we pumped the chamber for $\sim$ 24-48 hours in order to eliminate adsorbed molecules on the nanowire surfaces and the substrate. \cite{Gul}

\noindent
\textbf{Low-temperature measurement.}
All measurements were carried out in an Oxford $^{3}$He/$^{4}$He-dilution refrigerator equipped with a superconducting magnet, using a standard four-terminal measurement configuration. The current and voltage preamplifiers were calibrated by standard resistors. In order to minimise the electronic noise, we used a series of $\pi$-filters, copper-powder filter and RC filters at different temperature stages for covering different frequency ranges\cite{Xiang, Jarillo}. Whenever needed, a magnetic field was applied perpendicularly to the substrate plane and thus to the nanowires in this work.

\begin{center}\noindent{\bf References}\end{center}


\newpage
\noindent
\textbf{Acknowledgments}

\noindent
We acknowledge M. T. Deng and C. L. Yu for technical support in device fabrication, and L. Lu, J. Wei, and X. Lin for helpful discussions. This work was financially supported by the National Basic Research Program of the Ministry of Science and Technology of China (Grant No. 2012CB932703 and 2012CB932700), and by the National Natural Science Foundation of China (Grant No. 11374019, 91221202, 91421303, and 61321001). N.K. thanks the Ph.D. Program Foundation of the Ministry of Education of China for financial support (Grant No. 20120001120126). H.Q.X. also acknowledges financial support from the Swedish Research Council (VR).\\

\noindent
\textbf{Author Contributions}

\noindent
N.K. and H.Q.X. conceived and designed the experiments. P.C. developed and conducted the growth of the nanowires. S.L. and D.X.F. fabricated the devices. S.L., D.X.F. and N.K. performed the electrical measurements. L.B.W. and Y.Q.H. helped with low-temperature measurements. S.L. and N.K. analyzed the measurement data. S.L., N.K. and H.Q.X. co-wrote the manuscript. H.Q.X. supervised the project. All authors discussed the results and commented on the manuscript.\\

\noindent
\textbf{Additional information}

\noindent
\textbf{Competing interests}
\noindent
The authors declare no competing financial interests.

\newpage

\begin{figure}
\begin{center}
\includegraphics[width=5.8in]{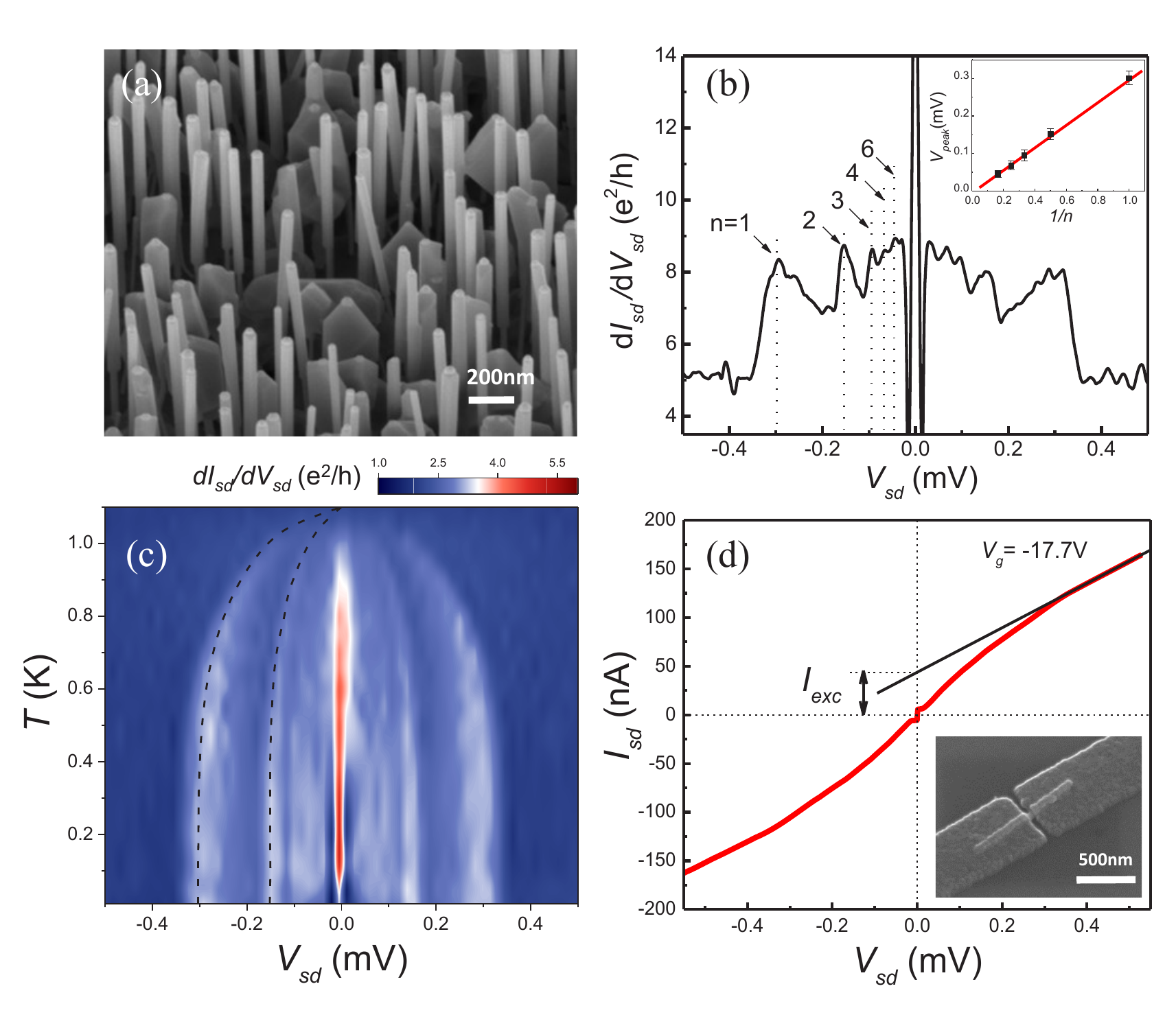}
\caption{\textbf{Multiple Andreev reflections in an InSb nanowire Josephson junction device.} \textbf{a}, SEM image of as-grown InSb nanowires grown via an InP/InAs stem technique on an InP(111)B substrate. These nanowires are 50-90 nm in diameter. Gold seed particles are visible on top of individual nanowires.
\textbf{b}, Differential conductance $dI_{sd}/dV_{sd}$ versus bias voltage $V_{sd}$ measured for device D1 made from an InSb nanowire with a diameter of $\sim$56 nm at gate voltage $V_{g} = -15.25$ V and base temperature $T$ = 10 mK. The arrows and vertical dotted lines indicate conductance peaks at the voltage positions of $eV_{n}=2\Delta/n$ ($n$=1,2,...), corresponding to multiple Andreev reflection processes. The inset shows the peak position versus the inverse of the peak index $1/n$. The data points are averaged for several  $dI_{sd}/dV_{sd}$ traces of different gate voltages and the error bars denote the standard deviation. The solid line is a linear fit to the peak positions and is seen to go through the origin as expected.
\textbf{c}, $dI_{sd}/dV_{sd}$ as a function of $V_{sd}$ and temperature $T$ at zero magnetic field. The dashed lines follow the temperature evolutions of the conductance peaks due to multiple Andreev reflection processes, as expected from the temperature dependence of $\Delta(T)$ predicted by the BCS theory. The high zero-bias conductance peak originates from the proximity induced superconductivity in the nanowire.
\textbf{d}, Current-voltage characteristics measured at $V_{g} = -17.7$ V and $T$ = 10 mK. The linear fit of the current-voltage curve at high bias voltages  ($V_{sd}>2\Delta/e$) extrapolates to a finite value of the current at zero bias voltage, i.e., excess current $I_{exc}$.
The inset shows an SEM image of the corresponding InSb nanowire-based Josephson junction. Here, the separation between the two contacts is $\sim$ 60 nm.}
\label{fig1}
\end{center}
\end{figure}
\newpage

\begin{figure}
\begin{center}
\includegraphics[width=6.0in]{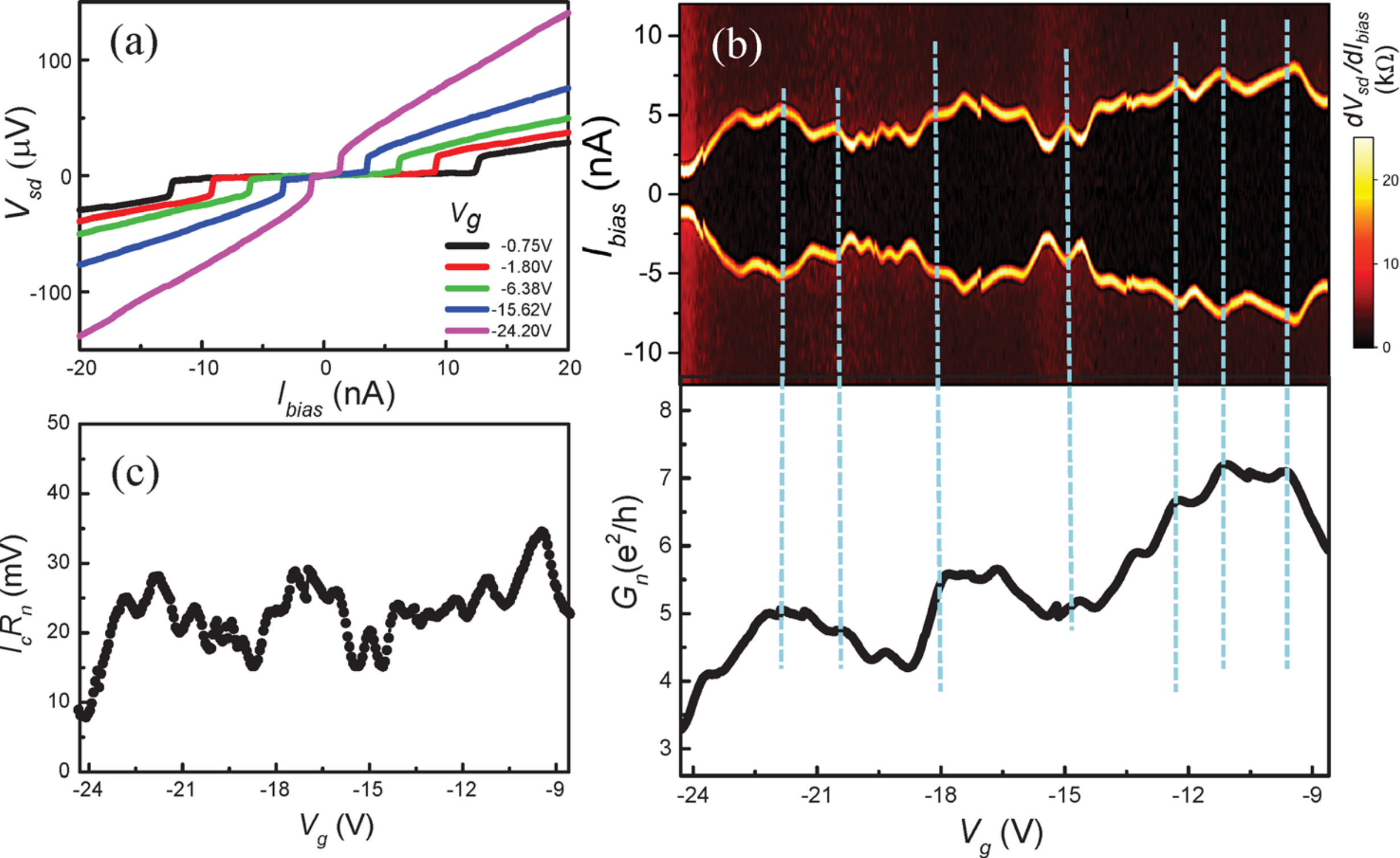}
\caption{\textbf{Gate-modulated supercurrent in an InSb nanowire Josephsen junction.} \textbf{a}, Measured voltage $V_{sd}$ as a function of bias current $I_{bias}$ for device D1 at $T$ = 10 mK and different gate voltages $V_{g}$. Here a gate-tunable supercurrent is seen.
\textbf{b}, Upper panel: Differential resistance $dV_{sd}/dI_{bias}$ as a function of $I_{bias}$ and $V_{g}$. The central dark area is the region with $dV_{sd}/dI_{bias}=$ 0, i.e., the device in the superconducting state. The bright boundaries correspond to the transitions from the superconducting state to the dissipative state.
The critical current $I_{c}$, measured by the bright boundaries, oscillates as a function of $V_g$.
Lower panel: Normal state conductance $G_{n}$ of device D1 extracted from linear fits to the $I_{sd}-V_{sd}$ curves at the high bias voltage region with $V_{sd} > 2\Delta/e$. Here, the correlations between the modulations of the critical current and the oscillations of the normal state conductance are clearly seen as indicated by vertical dashed lines.
\textbf{c}, Product of the critical current $I_{c}$ and the normal state resistance $R_{n}$ of the device extracted as a function of $V_g$. } \label{fig2}
\end{center}
\end{figure}
\newpage

\begin{figure}
\begin{center}
\includegraphics[width=6.0in]{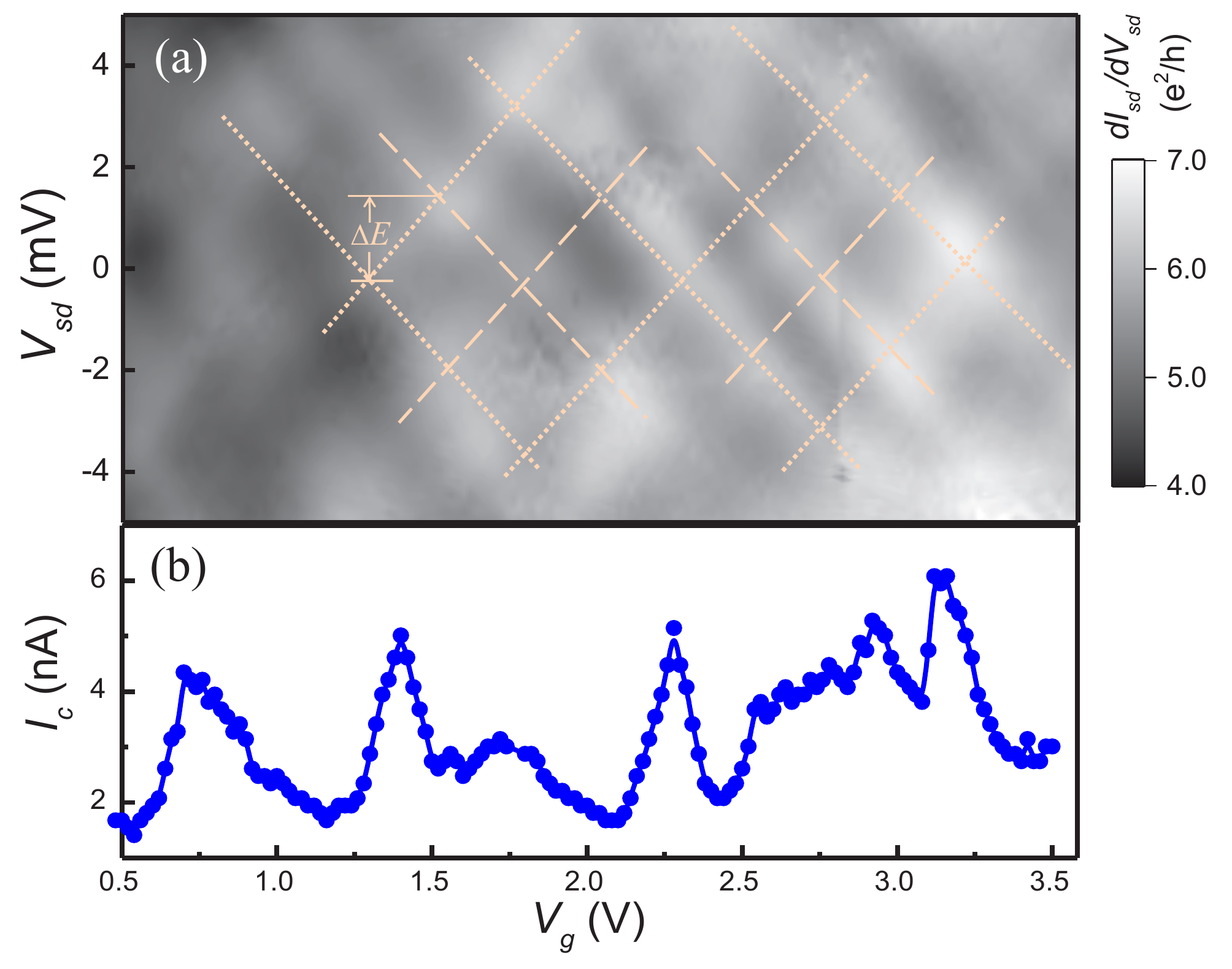}
\caption{\textbf{Fabry-P\'{e}rot interference in a ballistic InSb nanowire Josephson junction. } \textbf{a}, Normal-state differential conductance $dI_{sd}/dV_{sd}$ measured for device D2 with $L\sim 95$ nm in contact separation and $D\sim 60$ nm in nanowire diameter as a function of source-drain voltage $V_{sd}$ and gate voltage $V_{g}$ at $T$ = 10 mK and a magnetic field of 50 mT. Here, a Fabry-P\'{e}rot-like interference pattern is observable, as indicated by dashed and dotted lines in the figure. The smallest energy spacing of the Fabry-P\'{e}rot resonances is marked as $\Delta E$ in the plot.
\textbf{b}, Critical current $I_c$ of the device measured in the same gate voltage range as in  \textbf{a}.} \label{fig3}
\end{center}
\end{figure}
\newpage

\begin{figure}
\begin{center}
\includegraphics[width=6.2in]{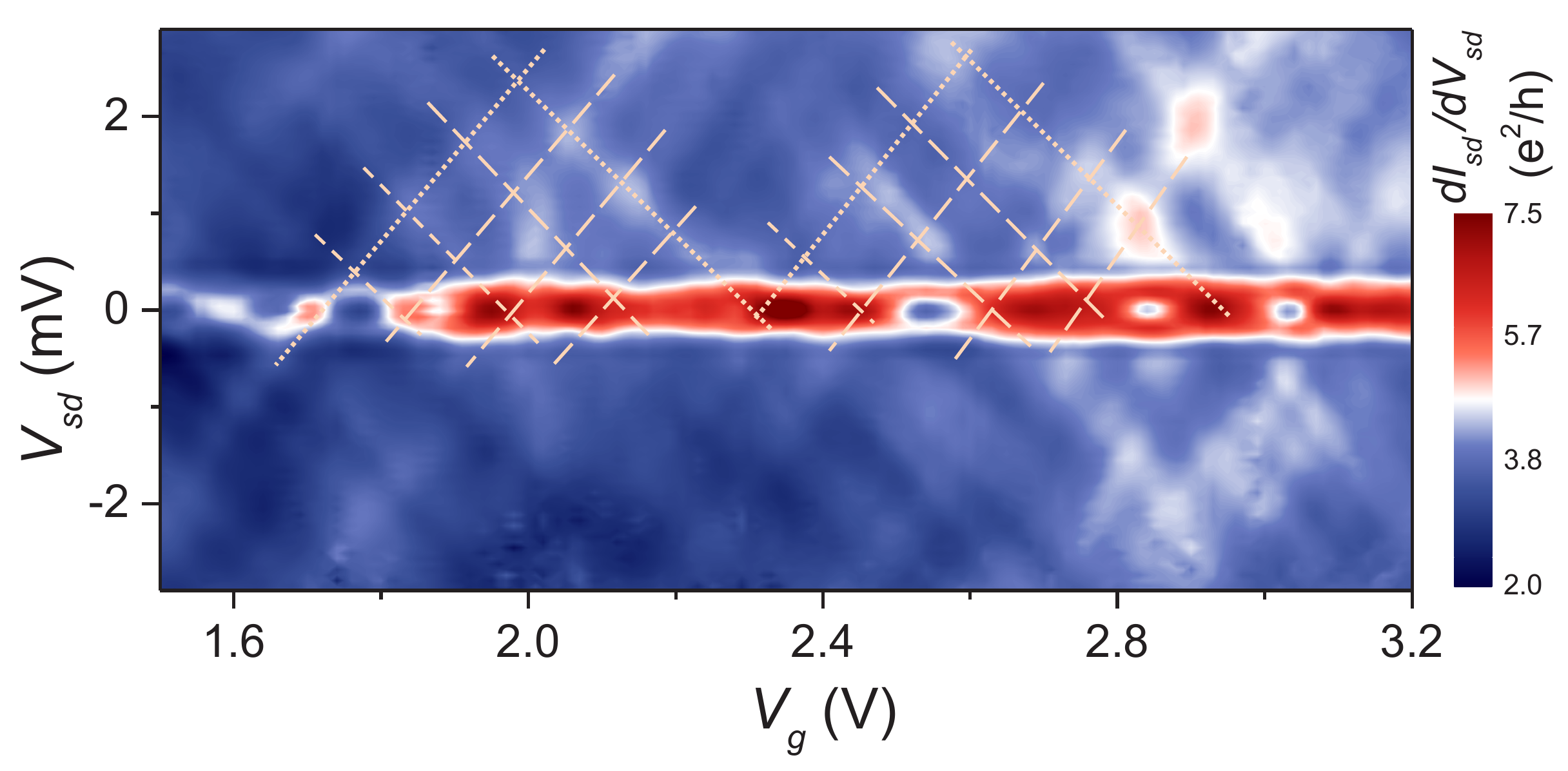}
\caption{\textbf{Charge stability diagram of an InSb nanowire Josephson junction at $B=0$.} Differential conductance $dI_{sd}/dV_{sd}$ versus $V_{sd}$ and $V_{g}$ measured for device D3 with $L\sim 65$ nm in contact separation and $D\sim 90$ nm in nanowire diameter at zero magnetic field and $T$ = 10 mK.
At bias voltages below $2\Delta/e$, the differential conductance oscillations quasi-periodically, indicating that the supercurrent is modulated, as a result of the formation of resonant states in the nanowire junction by Fabry-P\'{e}rot interference.
The dashed and dotted lines in the figure are guides to the eyes, highlighting the checkerboard pattern of Fabry-P\'{e}rot conductance oscillations at high bias voltages.} \label{fig4}
\end{center}
\end{figure}


\begin{thebibliography}{99}

\bibitem{SNS} De Franceschi, S., Kouwenhoven, L. P., Sch\"{o}nenberger, C. \&  Wernsdorfer, W. Hybrid superconductor-quantum dot devices. {\em Nat. Nanotechnol.} {\bf 5,} 703-711 (2010).

\bibitem{Yeyati} Mart\'{i}n-Rodero, A. \& Levy Yeyati, A. Josephson and Andreev transport through quantum dots. {\em Adv. Phys.} {\bf 60,} 899-958 (2011).

\bibitem{Doh} Doh, Y. -J. {\em et al.} Tunable Supercurrent through semiconductor nanowires. {\em Science} {\bf 309,} 272-275 (2005).

\bibitem{Xiang} Xiang, J., Vidan, A., Tinkham, M., Westervelt, R. M. \& Lieber, C. M. Ge/Si nanowire mesoscopic Josephson junctions. {\em Nat. Nanotechnol.} {\bf 1,} 208-213 (2006).

\bibitem{Jarillo} Jarillo-Herrero, P., van Dam, J. A. \& Kouwenhoven, L. P. Quantum supercurrent transistors in carbon nanotubes. {\em Nature} {\bf 439,} 953-956 (2006).

\bibitem{SiGe} Katsaros, G. {\em et al.} Hybrid superconductor-semiconductor devices made from self-assembled SiGe nanocrystals on silicon. {\em Nat. Nanotechnol.} {\bf 5,} 458-464 (2010).

\bibitem{C60} Winkelmann, C. B., Roch, N., Wernsdorfer, W., Bouchiat, V. \& Balestro, F. Superconductivity in a single-C60 transistor. {\em Nat. Phys.} {\bf 5,} 876-879 (2009).

\bibitem{CEA} Pillet, J.-D. {\em et al.} Andreev bound states in supercurrent-carrying carbon nanotubes revealed. {\em Nat. Phys.} {\bf 6,} 965-969 (2010).

\bibitem{Lee} Lee, E. J. H. {\em et al.} Spin-resolved Andreev levels and parity crossings in hybrid superconductor-semiconductor nanostructures. {\em Nat. Nanotechnol.} {\bf 9,} 79-84 (2014).

\bibitem{Basel} Hofstetter, L., Csonka, S., Nygard, J. \& Sch\"onenberger, C. Cooper pair splitter realized in a two-quantum-dot Y-junction. {\em Nature} {\bf 461,} 960-963 (2009).

\bibitem{Strunk} Herrmann, L. G. {\em et al.} Carbon nanotubes as cooper-pair beam splitters. {\em Phys. Rev. Lett.} {\bf 104,} 026801 (2010).

\bibitem{Giazotto} Giazotto, F., Heikkil\"a, T. T., Luukanen, A., Savin, A. M. \& Pekola, J. P. Opportunities for mesoscopics in thermometry and refrigeration: Physics and applications. {\em Rev. Mod. Phys.} {\bf 78,} 217-274 (2006).

\bibitem{Kane} Fu, L. \& Kane, C. L. Superconducting proximity effect and Majorana fermions at the surface of a topological insulator. {\em Phys. Rev. Lett.} {\bf 100,} 96407 (2008).

\bibitem{Lutchyn01} Lutchyn, R. M., Sau, J. D. \& Das Sarma, S. Majorana fermions and a topological phase transition in semiconductor-superconductor heterostructures. {\em Phys. Rev. Lett.} {\bf 105,} 77001 (2010).

\bibitem{Oreg} Oreg, Y., Refael, G. \& von Oppen, F. Helical liquids and Majorana bound states in quantum wires. {\em Phys. Rev. Lett.} {\bf 105,} 177002 (2010).

\bibitem{Nilsson} Nilsson, H. A., Samuelsson, P., Caroff, P. \& Xu, H. Q. Supercurrent and multiple Andreev reflections in an InSb nanowire Josephson junction. {\em Nano Lett.} {\bf 12,} 228-233 (2012).

\bibitem{Plissard2013} Plissard, S. R. {\em et al.} Formation and electronic properties of InSb nanocrosses. {\em Nat. Nanotechnol.} {\bf 8,} 859-864 (2013).

\bibitem{Mourik} Mourik, V. {\em et al.} Signatures of Majorana fermions in hybrid superconductor-semiconductor nanowire devices. {\em Science} {\bf 336,} 1003-1007 (2012).

\bibitem{Deng} Deng, M. T. {\em et al.} Anomalous zero-bias conductance peak in a Nb-InSb nanowire-Nb hybrid device. {\em Nano Lett.} {\bf 12,} 6414-6419 (2012).

\bibitem{Churchill} Churchill H. O. H. {\em et al.} Superconductor-nanowire devices from tunneling to the multichannel regime: Zero-bias oscillations and magnetoconductance crossover. {\em Phys. Rev. B} {\bf 87,} 241401(R) (2013)

\bibitem{Alicea} Alicea, J., Oreg, Y., Refael, G., von Oppen, F. \& Fisher, M. P. A. Non-Abelian statistics and topological quantum information processing in 1D wire networks. {\em Nat. Phys.} {\bf 7,} 412-417 (2011).

\bibitem{Heck} van Heck, B., Akhmerov, A. R., Hassler, F., Burrello, M. \& Beenakker, C. W. J. Coulomb-assisted braiding of Majorana fermions in a Josephson junction array. {\em New J. Phys.} {\bf 14,} 035019 (2012).

\bibitem{Sau} Sau, J. D., Clarke, D. J. \& Tewari, S., Controlling non-Abelian statistics of Majorana fermions in semiconductor nanowires. {\em Phys. Rev. B} {\bf 84,} 094505 (2011).

\bibitem{Fu} Fu, L. \& Kane, C. L. Josephson current and noise at a superconductor/quantum-spin-Hall-insulator/superconductor junction. {\em Phys. Rev. B} {\bf 79,} 161408 (2009).

\bibitem{Cheng} Cheng, M. \& Lutchyn, R. M. Josephson current through a superconductor/semiconductor-nanowire/superconductor junction: Effects of strong spin-orbit coupling and Zeeman splitting. {\em Phys. Rev. B} {\bf 86,} 134522 (2012).

\bibitem{Nazarov1} Yokoyama, T., Eto, M. \& Nazarov, Yu. V. Josephson current through semiconductor nanowire with spin-orbit interaction in magnetic field. {\em J. Phys. Soc. Jpn.} {\bf 82,} 054703 (2013).

\bibitem{Cayao} Cayao, J., Prada, E., San-Jose, P. \& Aguado, R. SNS junctions in nanowires with spin-orbit coupling: Role of confinement and helicity on the subgap spectrum. {\em Phys. Rev. B} {\bf 91,} 024514 (2015).

\bibitem{XuNanotech2012} Xu, T. {\em et al.} Faceting, composition and crystal phase evolution in III-V antimonide nanowire heterostructures revealed by combining microscopy techniques. {\em Nanotechnology} {\bf 23,} 095702 (2012).

\bibitem{Thelander2012} Thelander, C., Caroff, P., Pissard, S. \& Dick, K. A. Electrical properties of InAs$_{1-x}$Sb$_{x}$ and InSb nanowires grown by molecular beam epitaxy. {\em Appl. Phys. Lett.} {\bf 100,} 232105 (2012).

\bibitem{Fan} Fan, D. {\em et al.} Formation of long single quantum dots in high quality InSb nanowires grown by molecular beam epitaxy. {\em Nanoscale} {\bf 7,} 14822-14828 (2015)

\bibitem{Liu} Liu J., Potter A. C., Law K. T. \& Lee P. A. Zero-Bias Peaks in the Tunneling Conductance of Spin-Orbit Coupled Superconducting Wires with and without Majorana End States. {\em Phys. Rev. Lett.} {\bf 109,} 267002 (2012)

\bibitem{BTK} Blonder, G. E., Tinkham, M. \& Klapwijk, T. M. Transition from metallic to tunneling regimes in superconducting microconstrictions: Excess current, charge imbalance, and supercurrent conversion. {\em Phys. Rev. B} {\bf 25,} 4515-4532 (1982).

\bibitem{Elke} Scheer, E. {\em et al.} The signature of chemical valence in the electrical conduction through a single-atom contact. {\em Nature} {\bf 394,} 154-157 (1998).

\bibitem{Cuevas} Cuevas, J. C., Mart\'{i}n-Rodero, A. \& Levy Yeyati, A. Hamiltonian approach to the transport properties of superconducting quantum point contacts. {\em Phys. Rev. B} {\bf 54,} 7366-7379 (1996).

\bibitem{Flensberg} Flensberg, K., Hansen, J. B. \& Octavio, M. Subharmonic energy-gap structure in superconducting weak links. {\em Phys. Rev. B} {\bf 38,} 8707-8711 (1988).

\bibitem{OBTK} Octavio, M., Tinkham, M., Blonder, G. E. \& Klapwijk, T. M. Subharmonic energy-gap structure in superconducting constrictions. {\em Phys. Rev. B} {\bf 27,} 6739-6746 (1983).

\bibitem{Likharev} Likharev, K. K. Superconducting weak links. {\em Rev. Mod. Phys.} {\bf 51,} 101-159 (1979).

\bibitem{Beenakker02} Beenakker, C. W. J. Three Universal Mesoscopic Josephson Effects in {\em Proceedings of the 14th Taniguchi International Symposium on Physics of Mesoscopic Systems} (eds Fukuyama, H. and Ando T.) 235-254 (Springer, Berlin, 1992).

\bibitem{Tinkham} Tinkham, M. {\em Introduction to Superconductivity.} (Dover, Mineola, NY, 1996).

\bibitem{Nitta} Takayanagi, H., Hansen, J. B. \& Nitta, J. Localization effects on the critical current of a superconductor-normal metal-superconductor junction {\em Phys. Rev. Lett.} {\bf 74,} 162 (1994).

\bibitem{Abay} Abay, S. {\em et al.} Quantized conductance and its correlation to the supercurrent in a nanowire connected to superconductors. {\em Nano Lett.} {\bf 13,} 3614 (2013).

\bibitem{Lindelof} J{\o}rgensen, H. I., Grove-Rasmussen, K., Novotn\'{y}, T., Flensberg, K. \& Lindelof, P. E. Electron transport in single-wall carbon nanotube weak links in the Fabry-Perot regime. {\em Phys. Rev. Lett.} {\bf 96,} 207003 (2006).

\bibitem{Lau} Liu, G., Zhang, Y. \& Lau, C. N. Gate-tunable dissipation and superconductor-insulator transition in carbon nanotube Josephson junctions. {\em Phys. Rev. Lett.} {\bf 102,} 016803 (2009).

\bibitem{Liang} Liang, W. {\em et al.} Fabry-Perot interference in a nanotube electron waveguide. {\em Nature} {\bf 411,} 665-669 (2001).

\bibitem{Javey} Javey, A., Guo, J., Wang, Q., Lundstrom, M. \& Dai, H. J. Ballistic carbon nanotube field-effect transistors. {\em Nature} {\bf 424,} 654 (2003).

\bibitem{Cao} Cao, J., Wang, Q. \& Dai, H. J. Electron transport in very clean, as-grown suspended carbon nanotubes. {\em Nat. Mater.} {\bf 4,} 745 (2005).

\bibitem{Kretinin} Kretinin, A. V., Popovitz-Biro, R., Mahalu, D. \& Shtrikman, H. Multimode Fabry-Perot conductance oscillations in suspended stacking-faults-free InAs nanowires {\em Nano Lett.} {\bf 10,} 3439 (2010).

\bibitem{Miao} Miao, F. {\em et al.} Phase-coherent transport in graphene quantum billiards. {\em Science} {\bf 317,} 1530-1533 (2007).

\bibitem{Hakonen} Oksanen, M. {\em et al.} Single-mode and multimode Fabry-Perot interference in suspended graphene. {\em Phys. Rev. B} {\bf 89,} 121414 (2014).

\bibitem{Glazman} Glazman, L. I. \& Matveev, K. A. Resonant Josephson current through Kondo impurities in a tunnel barrier. {\em JETP Lett.} {\bf 49,} 659-662 (1989).

\bibitem{Rittenhouse} Rittenhouse, G. E. \& Graybeal, J. M. Fabry-Perot interference peaks in the critical current for ballistic superconductor-normal-metal-superconductor Josephson junctions. {\em Phys. Rev. B} {\bf 49,} 1182 (1994).

\bibitem{Katine} Katine, J. A. {\em et al.}. Point Contact Conductance of an Open Resonator. {\em Phys. Rev. Lett.} {\bf 79,} 4806 (1997).

\bibitem{White} Gunlycke, D. \& White, C. T. Graphene interferometer. {\em Appl. Phys. Lett.} {\bf 93,} 122106 (2008).

\bibitem{Trauzettel} M\"{u}ller, M., Br\"{a}uninger, M. \& Trauzettel, B. Temperature dependence of the conductivity of ballistic graphene. {\em Phys. Rev. Lett.} {\bf 103,} 196801 (2009).

\bibitem{Zalipaev} Zalipaev V. V., Forrester D. M., Linton C. M. \& Kusmartsev F. V. Localised States of Fabry-Perot Type in Graphene Nano-Ribbons in {\em New Progress on Graphene Research} (ed Gong J. R.) Ch. 2, 29-79 (InTech, 2013)

\bibitem{CaroffSmall2008} Caroff, P. {\em et al.} High-quality InAs/InSb nanowire heterostructures grown by metal-organic vapor-phase epitaxy. {\em Small} {\bf 4,} 878-882 (2008).

\bibitem{ErcolaniNanotech2009} Ercolani, D. {\em et al.} InAs/InSb nanowire heterostructures grown by chemical beam epitaxy. {\em Nanotechnology} {\bf 20,} 505605 (2009).

\bibitem{PlissardNL2012} Plissard, S. R. {\em et al.} From InSb nanowires to nanocubes: looking for the sweet spot. {\em Nano Lett.} {\bf 12,} 1794 (2012).

\bibitem{Suyatin} Suyatin, D. B., Thelander, C., Bj\"{o}rk, M. T., Maximov, I. \& Samuelson, L. Sulfur passivation for ohmic contact formation to InAs nanowires. {\em Nanotechnology} {\bf 18,} 105307 (2007).

\bibitem{Gul} G\"{u}l, \"{O}., van Woerkon, D. J., van Weperen, I. Car, D., Plissard, S. R., Bakkers, E. P. A. M. \& Kouwenhoven, L. P. Towards high mobility InSb nanowire devices. {\em Nanotechnology} {\bf 26,} 215202 (2015).



\end{thebibliography}
\end{document}